\documentclass[prb,aps,10pt,twocolumn,superscriptaddress,longbibliography]{revtex4}
\usepackage{amsfonts,amsmath,bm}
\usepackage{color}
\usepackage{natbib}
\usepackage{hyperref}
\usepackage{physics}

\newcommand{\rr}{\bm{r}}
\usepackage{xcolor}
\usepackage[normalem]{ulem}

\usepackage{graphicx}
\newcommand{\blue}[1]{\textcolor{black}{#1}}

\begin{document}
	
	\title{Structural symmetry-breaking to explain radiative Auger transitions in self-assembled quantum dots}   
	
	\author{Krzysztof Gawarecki}
	\email{Krzysztof.Gawarecki@pwr.edu.pl}
	\affiliation{Department of Theoretical Physics, Wroc{\l}aw University of Science and Technology, 50-370 Wroc{\l}aw, Poland}
	\author{Clemens Spinnler}
	\affiliation{Department of Physics, University of Basel, Klingelbergstrasse 82, 4056 Basel, Switzerland}
	\author{Liang Zhai}
	\affiliation{Department of Physics, University of Basel, Klingelbergstrasse 82, 4056 Basel, Switzerland}
	\author{Giang N. Nguyen}
	\affiliation{Department of Physics, University of Basel, Klingelbergstrasse 82, 4056 Basel, Switzerland}
	\author{Arne Ludwig}
	\affiliation{Lehrstuhl f\"ur Angewandte Festk\"orperphysik, Ruhr-Universität Bochum, 44780 Bochum, Germany}
	\author{Richard J. Warburton}
	\affiliation{Department of Physics, University of Basel, Klingelbergstrasse 82, 4056 Basel, Switzerland}
	\author{Matthias C. L\"obl\footnote{Current address: Center for Hybrid Quantum Networks (Hy-Q), The Niels Bohr Institute, University of Copenhagen, DK-2100 Copenhagen Ø, Denmark}}
	\affiliation{Department of Physics, University of Basel, Klingelbergstrasse 82, 4056 Basel, Switzerland}
	\author{Doris E. Reiter}
	\affiliation{Institut f{\"u}r Festk{\"o}rpertheorie, Universit{\"a}t M{\"u}nster, 48149 M{\"u}nster, Germany}
	\author{Pawe{\l} Machnikowski}
	\affiliation{Department of Theoretical Physics, Wroc{\l}aw University of Science and Technology, 50-370 Wroc{\l}aw, Poland}
	
	\begin{abstract}
		The optical spectrum of a quantum dot is typically dominated by the fundamental transition between the lowest-energy configurations. However, the radiative Auger process can result in additional red-shifted emission lines. The origin of these lines is a combination of Coulomb interaction and symmetry-breaking in the quantum dot. In this paper, we present measurements of such radiative Auger lines for a range of InGaAs/GaAs self-assembled quantum dots and use a tight-binding model with a configuration interaction approach to explain their appearance. Introducing a composition fluctuation cluster in the dot, our calculations show excellent agreement with measurements. We relate our findings to group theory explaining the origin of the additional emission lines. Our model and results give insight into the interplay between the symmetry breaking in a quantum dot and the position and strength of the radiative Auger lines.  
	\end{abstract}
	
	\maketitle
	
	\section{Introduction}
	\label{sec:intro}
	A self-assembled quantum dot (QD) can often be treated as a few-level system, in the simplest case as a two-level system, with discrete transition lines in the optical spectrum. Most pronounced is the fundamental transition associated with the recombination of the ground-state electron-hole pair. Such a transition can result in the emission of single-photons to be used for quantum technologies \cite{Santori2002,Schweickert2018,Wang2020mic,Tomm2021,Zhai2022}.
	
	Recently, additional spectral lines red-shifted from the fundamental trion transition have been observed on a single negatively charged QD \cite{Lobl2020}, as seen in Fig.~\ref{fig:spectrum-exp} with the details being explained in Sec.~\ref{sec:experiment}. Remarkably, it is possible to optically drive these transitions \cite{Spinnler2021}. In a singly charged QD, the fundamental transition is associated with an electron-hole recombination that originates from the lowest-energy three-particle (trion) state and leaves the additional (`spectator') single electron or hole in its ground state. The additional lines stem from radiative Auger transitions mediated by Coulomb interaction \cite{Lobl2020}. In the radiative Auger process the energy of the recombining electron-hole pair is partially transferred to an intraband electronic excitation, which promotes the resident carrier to an excited state and red-shifts the photon emission. Analogous transitions have been observed in the X-ray spectra of atoms \cite{Bloch1935,Bloch1935a,Aberg1969,Aberg1971} and, more recently, in optical spectra of semiconductor nanostructures \cite{Antolinez2019,Jordi20}. Note that this effect includes photons and therefore differs from the Auger scattering between electrons only, typically resulting in electron emission from the QD \cite{lochner2020real,mannel2021auger}. When explored in QDs, the radiative Auger lines can be used for characterization of otherwise unreachable single-particle excitation energies and provide temporal characteristics of single-carrier relaxation \cite{Lobl2020,Spinnler2021}.
	
	To describe theoretically the radiative Auger transitions, models including higher excited QD states accounting for the Coulomb mixing between these states need to be employed \cite{Gaweczyk2017,Holtkemper2018,huber2019single,Zielinski2020}. In a perfectly symmetric QD \blue{(and neglecting the atomic disorder)}, most of these transitions would be forbidden by symmetry. Accordingly, in order to account for the radiative Auger lines it is important to consider asymmetries and imperfections. In addition, some observed features are not straightforward to explain. In particular, the strength of the Auger lines varies from QD to QD and may reach values on the order of one percent of the fundamental line \cite{Lobl2020} and the observed transitions do not show spin-related Zeeman splittings in a magnetic field, indicating a unique final spin state. 
	
	In this paper, we provide a thorough understanding of the radiative Auger process in QDs including the above mentioned peculiarities. We consider measurements on four InGaAs/GaAs QDs and model them using a tight binding model. With this, we explain the relatively strong brightness of the radiative Auger lines by lowering of the QD symmetry on the structural level. By assuming compositional inhomogeneity in the form of an indium-rich cluster within the QD volume, we reproduce the experimentally observed transition strengths. Using group theory we explain the emerging of the different lines by invoking symmetry breaking on the level of the QD shape. 
	

	\section{The radiative Auger transitions}
	\label{sec:experiment}
	
	\begin{figure}[tb]
		\begin{center}
			\includegraphics[width=\columnwidth]{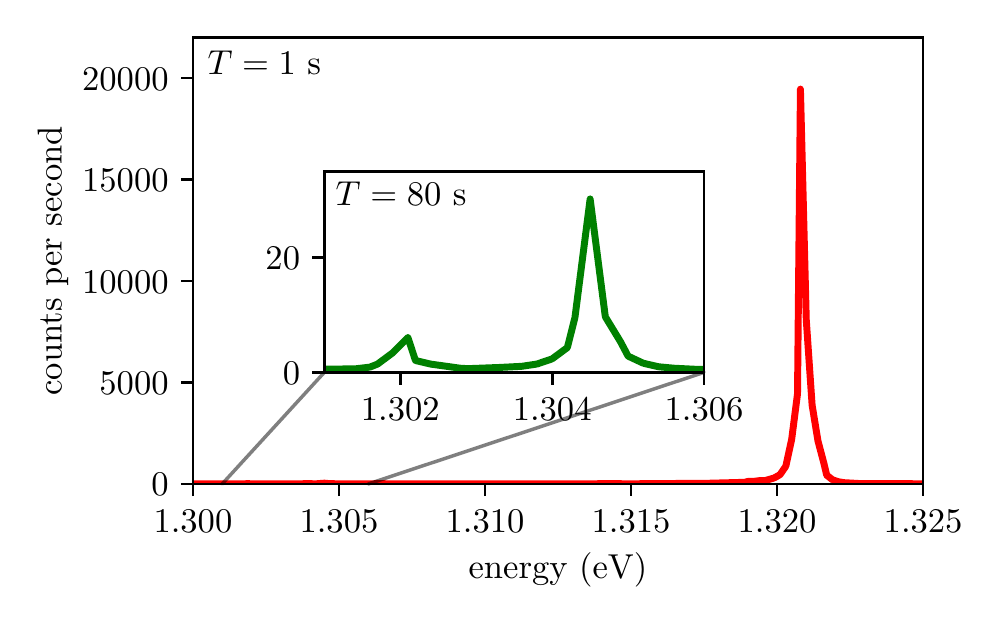}
		\end{center}
		\caption{\label{fig:spectrum-exp} Measured resonance fluorescence spectrum of an InGaAs/GaAs QD integrated for 1~s (red line, main panel) and 80~s (green line, inset). The spectrum corresponds to QD1 in Fig.~\ref{fig:bfield}.}
	\end{figure}
	
	Before introducing the theoretical model, we briefly account for the experimental observation and the mechanism of the radiative Auger transitions in QDs. In the simplest, highly symmetric single-particle model one obtains a series of transitions between valence and conduction band. For each transition line the initial and final state belong to the same single-particle shell of the QD spectrum. Additionally, the created electron and hole have opposite projections of the envelope angular momentum, corresponding to null envelope angular momentum projection of the electron-hole pair \cite{jacak98a}. The resulting shell structure of optical excitations, corresponding to the sequence of $s$--$s$, $p$--$p$ etc. transitions, has indeed been observed experimentally \cite{hawrylak00,Raymond2004}.
	
	Coulomb interaction modifies this simple picture described above in several ways. It renormalizes the energy of the electron-hole pair (which becomes an exciton) and introduces energy shifts between transitions taking place in the presence of other carriers (biexciton or trion transitions) \cite{wojs95,Hawrylak2003}. It splits the lines according to their spin configurations due to exchange interactions \cite{bayer02b}. It also hybridises two-particle configurations \cite{Ardelt2016b}. The latter effect has been demonstrated in a double-QD structure, where the symmetry is lowered by a lateral offset of the QDs. The configuration mixing may then involve two-particle states of different angular momentum projections, which is revealed in the optical spectra of the system \cite{Ardelt2016b}. The radiative Auger process is another consequence of the Coulomb interaction and reveals Coulomb-induced configuration mixing in a single QD.
	\begin{figure}[tb]
		\begin{center}
			\includegraphics[width=\columnwidth]{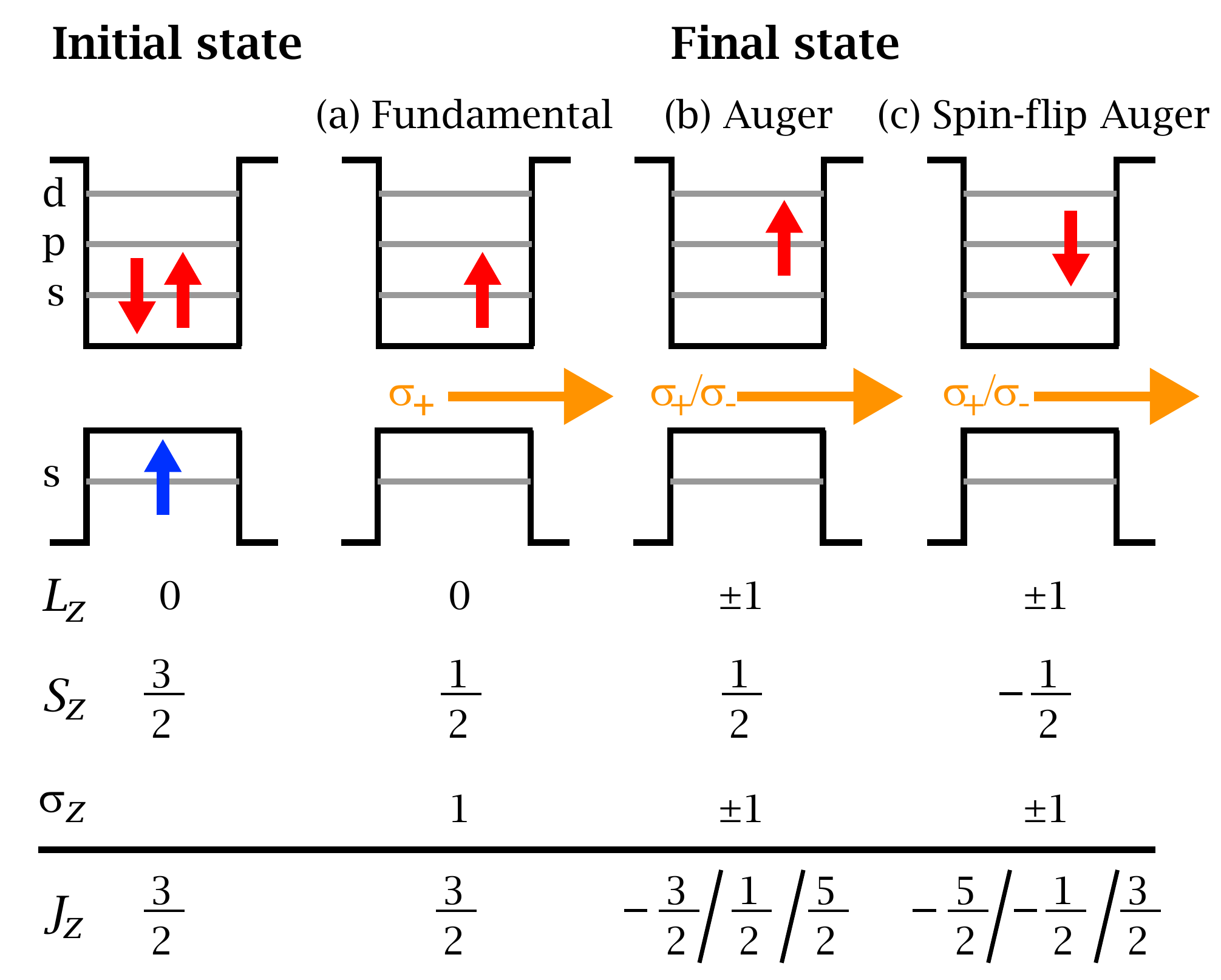}
		\end{center}
		\caption{\label{fig:SO-diagram}Angular momentum in ground-state trion recombination: The initial state (left) decays into one of the final states yielding either (a) the fundamental transition, (b) the Auger transition to the $p$-shell or (c) the spin-flip Auger transition to the $p$-shell. Below each diagram we account for the angular momentum of the state, with $L_z$ and $S_z$ denoting the $z$ projections of the envelope and band angular momenta of the carrier(s), respectively, $\sigma_z$ representing the photon angular momentum \blue{(and the $\sigma_\pm$ denotes right and left circularly polarized photons)}, and $J_z$ denotes the total angular momentum.
		}
	\end{figure}

	The structures investigated here are InGaAs QDs grown in the Stranski-Krastanov mode and using a flushing technique \cite{Wasilewski1999,Ludwig2017}. For controlling the charge state of the QDs, they are placed in an n-i-p diode with a tunnel barrier of $40$~nm between QDs and a n-type (silicon doped) backgate. 
	The spectrum of scattered light (resonance fluorescence) was recorded upon resonant excitation of a single negatively charged QD \cite{Lobl2020}. At non-zero magnetic fields a selected spin configuration of the trion was excited by appropriate tuning of the resonant excitation.
	All experiments are performed at 4.2~K using a dark-field microscope that can distinguish QD emission from back-reflected laser light by a cross-polarization scheme.
	Details on fabrication and measurement of the QDs can be found in Ref.~\cite{Lobl2020}.
	
	An example of the observed spectra is shown in Fig.~\ref{fig:spectrum-exp}. \blue{Some general properties of the measured spectra can be described using the idealized diagrams shown in Fig.~\ref{fig:SO-diagram}, where we characterize the initial and final states using
		the spin, band and envelope angular momenta (despite the fact that in presence of spin-orbit interaction and symmetry breaking, they are not strictly good quantum numbers).} The first measurement \blue{of Fig.~\ref{fig:spectrum-exp}} (red line), performed using the integration time of 1~s, reveals a strong line at about 1.321~eV. It corresponds to the fundamental trion transition in which the `spectator' electron remains in its ground state. 
	This situation is depicted schematically in Fig.~\ref{fig:SO-diagram}, where the initial state is represented on the left, followed by the final state after optical emission [(a) for the fundamental transition]. Here the depicted levels and their labels ($s,p,d$) represent shells corresponding to the envelope states, while the Bloch wave functions are assumed $p$-type and $s$-type for the valence and conduction bands, respectively.  We stress that the fundamental transition is expected in the naive picture of uncorrelated interband transitions in a charged QD, where the band angular momentum changes by one and the symmetry of the envelope function is conserved. Below the diagrammatic figures, we indicate the $z$-projections of the envelope angular momentum ($L_z$), the band angular momentum ($S_z$) as well as the total angular momentum ($J_z$). Because the photon carries away an angular momentum of $\sigma_z$, the fundamental transition conserves the projection of the angular momentum on the symmetry axis and is therefore allowed. 
	
	While the strong fundamental transition line appears to be the only feature at short integration times, two additional lines appear at lower energies in the measurement performed with extended integration time, plotted in the inset to Fig.~\ref{fig:spectrum-exp}. For this particular QD, the relative intensity (compared to the fundamental transition) of these two lines is $0.34\cdot 10^{-3}$ and $1.2\cdot 10^{-3}$, respectively. Note that the overall intensities of the Auger lines as well as the ratio of the two Auger intensities strongly varies between different QDs as discussed in detail in Sec.~\ref{sec:results1}. These lines are attributed to radiative Auger transitions. In the \blue{idealized} picture of the shells, this corresponds to the recombination of one electron-hole pair, while the remaining electron is promoted to the $p$-shell. This is schematically shown in Fig.~\ref{fig:SO-diagram}(b,c), where we indicate the two possible orientations of the electron spin in the final state. Note that, apart from the spin degeneracy, the $p$-shell in Fig.~\ref{fig:SO-diagram} is further two-fold degenerate with respect to the envelope angular momentum, while in realistic QDs this degeneracy is lifted by anisotropy \blue{(which is caused by the inversion asymmetry of the underlying lattice, enhanced by anisotropy in the QD shape, atomic disorder, composition profile, strain, piezoelectric field, or magnetic field)}, yielding the two Auger lines visible in Fig.~\ref{fig:spectrum-exp}. At the bottom of Fig.~\ref{fig:SO-diagram}(b,c) we list the values of the projection of the angular momentum in the initial state and in the final states. Note that the spin-flip Auger transition (c) can conserve angular momentum while the transition (b) cannot. 
	
	To explain the observed features as well as the relative intensities of the Auger lines, we develop a theoretical model that connects the observations to symmetry considerations.

	\section{Theoretical model}
	\label{sec:model}
	
	\subsection{QD shape and parameters} \label{sec:shape}
	\begin{figure}[tb]
		\begin{center}
			\includegraphics[width=\columnwidth]{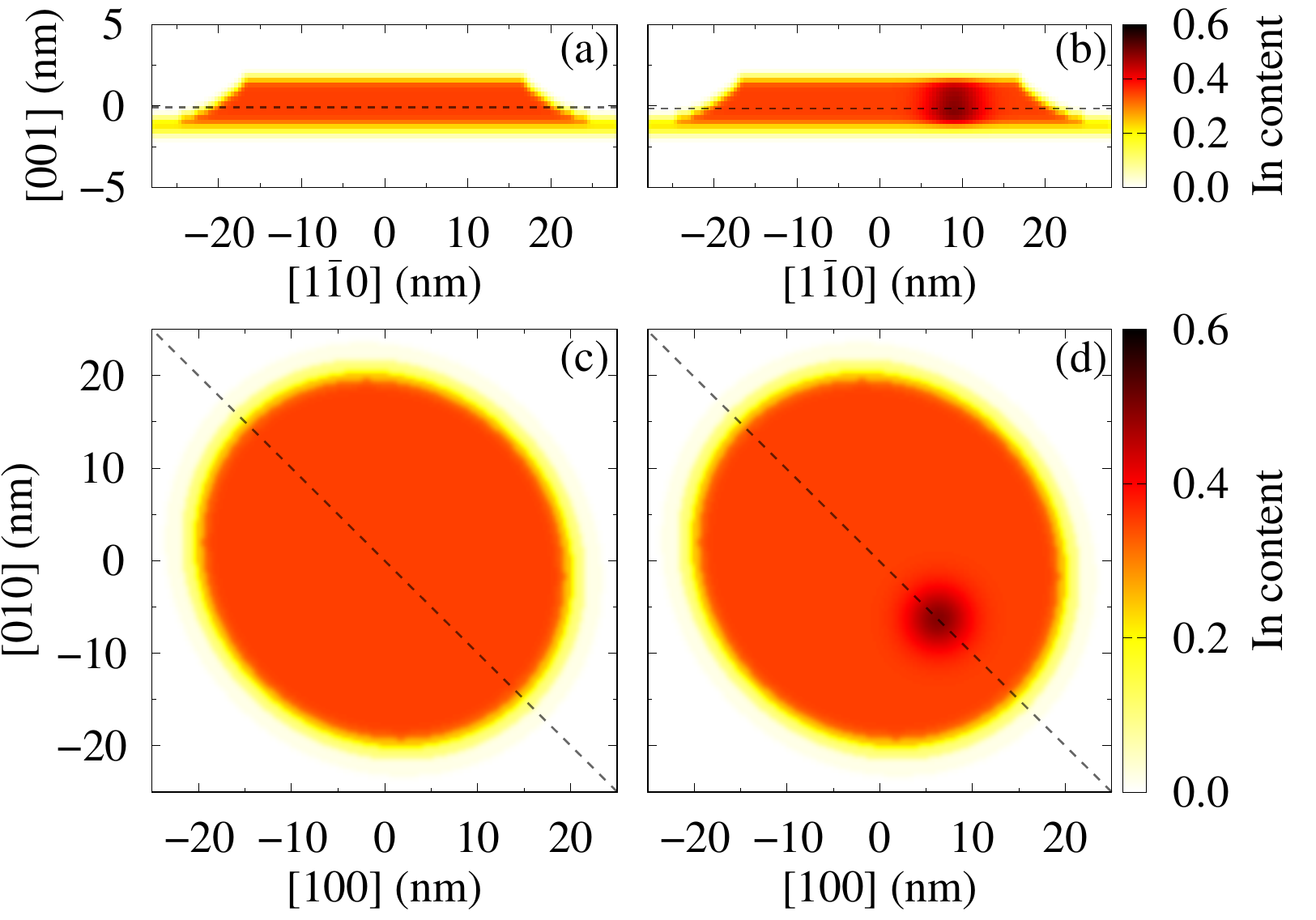}
		\end{center}
		\caption{\label{fig:comp} Cross-sections of compositions for InGaAs/GaAs QDs without (a,c) and with (b,d) composition cluster.}
	\end{figure}
	We assume an InGaAs QD in GaAs, which is elliptical in the lateral plane and capped in the z-plane as shown in Fig.~\ref{fig:comp}. Using \blue{a truncated elliptical}  Gaussian the surface of the dot is modelled by
	$$
	S= w \exp{-\frac{l^2 \qty( x \cos{\theta} + y \sin{\theta} )^2 + \qty( x \sin{\theta} - y \cos{\theta} )^2}{r^2_0}  },
	$$
	where $w$ is a parameter determining the slope, $2 r_0$ is related to the lateral extension, while $l$ describes the elongation in the direction given by $\theta$. At the height $h$ above the wetting layer the QD is truncated, which is consistent with the flushing step of the manufacturing process \cite{Wasilewski1999,Lobl2019CP}. 
	
	A further symmetry breaking is introduced by allowing for an additional cluster with a higher In content than assumed elsewhere in the QD, extending over a few lattice constants. An example is shown in Fig.~\ref{fig:comp}(b)/(d). The cluster is modeled by assuming the position-dependent composition profile of In$_{x(\rr)}$Ga$_{1-x(\rr)}$As with 
	$x(\rr) = c_0 + c(\rr)$, where $c_0$ is the nominal indium content in the QD and
	$$
	c(\rr) = C_{\mathrm{cl}} \exp{-\xi \qty[(x-x_0)^2 + (y-y_0)^2 + (z-z_0)^2]} \ 
	$$
	describes the locally increased indium content. Here $(x_0,y_0,z_0)$ and $\xi$ define the cluster position (counting from the bottom of the wetting layer) and its spatial extension, respectively, while $C_{\mathrm{cl}}$ denotes the maximum additional In content. \blue{As the model is atomistic, the composition distribution $x(\rr)$ refers to the probability of finding an In atom at given cation site. Such an approach results in inevitable symmetry breaking due to random compositional disorder.}
	
	We take the following parameters: The slope is $w = 25a$ (where \blue{$a \approx 0.565$~nm} is the GaAs lattice constant), $r_0 = 21a$, and the thickness of the wetting layer is chosen to be two monolayers corresponding to one lattice constant $a$. 
	The size and shape of the QD is taken to be similar to the one described in Ref.~\cite{Lobl2019CP}. Because InGaAs/GaAs QDs are often elongated in $[1\bar{1}0]$ direction~\cite{Joyce2001}, we include such an effect in our modeling. Therefore, we set the ellipticity to $l = (1.1)^{-1}$ and $\theta = -\pi/4$, that corresponds to $10\%$ elongation in the $[1\bar{1}0]$ direction. The composition is taken as In$_{0.35}$Ga$_{0.65}$As in the dot and In$_{0.2}$Ga$_{0.8}$As in the wetting layer. For the additional cluster we take $(x_0,y_0,z_0) = a(16/\sqrt{2},16/\sqrt{2},2)$, $\xi = 1/(6a)^2$. We consider two clusters: one with a higher In concentration, $C_{\mathrm{cl}} = 0.15$, and the other one with a lower In concentration, $C_{\mathrm{cl}} = 0.05$. Note that these assumed maximum additional indium contents are not very large compared to the overall nominal content of $c_0=0.35$. The composition distribution is processed by a Gaussian blur with a standard deviation of \blue{one GaAs lattice constant}. 
	
	\subsection{Coulomb interaction} \label{sec:coulomb}
	To calculate the Coulomb-coupled states, we use the tight-binding sp$^3$d$^5$s$^*$ implementation described in Refs.~\cite{Gawarecki2019b,Jancu1998} and utilize a configuration-interaction approach. The strain related to the lattice mismatch is accounted for within the valence force field model \cite{pryor98b}. The resulting strain-induced piezoelectric potential is calculated including the polarization up to the second order in strain tensor elements \blue{using} parameters from Ref.~\cite{bester06b}. 
	
	The wave functions of the single-particle states can be written as
	$$
	\ket{\Psi_{i}} = \sum_{n}^{N_\mathrm{a}} \sum_{\alpha}^{20} \varphi_{i,\alpha}(\bm{R}_{n}) \ket{\bm{R}_{n};\alpha},
	$$
	where $N_\mathrm{a}$ is the total number of atoms in the system, $\ket{\bm{R}_{n};\alpha}$ is an $\alpha$ atomic orbital on the site localized at $\bm{R}_{n}$, and $\varphi_{i,\alpha}(\bm{R}_{n})$ are complex coefficients.
	
	For the calculation of the Coulomb-coupled states, we change to the notation of the second quantization with $a^\dagger_{i}$ ($a_{i}$) and $h^\dagger_{i}$($h_{i}$) being the creation (annihilation) operators for the electron and hole single-particle states, respectively. With this, we calculate the negative trion states $\ket{X^-_\nu}$ consisting of two electrons and one hole. The corresponding Hamiltonian reads~\cite{Schulz2006,zielinski10}
	\begin{align}
		\label{eq:CI}
		H  =& \sum_{i}\epsilon^{(\mathrm{e})}_{i} a^{\dagger}_{i} a_{i} + \sum_{j} \epsilon^{(\mathrm{h})}_{j} h^{\dagger}_{j} h_{j} 
		+\frac{1}{2} \sum_{i i' j j'} V^{\mathrm{e e}}_{i j j' i'} a^{\dagger}_{i} a^{\dagger}_{j} a_{j'} a_{i'} \nonumber  \\
		& -\sum_{i i' j j'} V^{\mathrm{e h}}_{i j j' i'} a^{\dagger}_{i} h^{\dagger}_{j} h_{j'} a_{i'} 
		+ \sum_{i i' j j'} V^{\mathrm{eh, exch}}_{i j i' j'} a^{\dagger}_{i} h^{\dagger}_{j} a_{i'}  h_{j'}.
	\end{align}
	This Hamiltonian accounts for the electron/hole single particle energies via $\epsilon^{(\mathrm{e/h})}_{i}$ [first two terms in Eq.~\eqref{eq:CI}] as well as for the electron-electron and electron-hole Coulomb interaction (direct and exchange interaction) with the matrix elements given in Appendix. Because we only consider states with a single hole, the hole-hole Coulomb interaction does not contribute. We then use a configuration interaction approach to obtain the Coulomb-coupled trion states.
	
	The Hamiltonian is diagonalized in the basis of the lowest $n_\mathrm{e}$ electron and $n_\mathrm{h}$ hole states for expansion yielding the ground trion state $\ket{X^-}$ given by
	\begin{equation} \label{eq:trion}
		\ket{X^-} = \sum_{k}^{n_\mathrm{h}} \sum_{i,j}^{n_\mathrm{e}} c_{kij} \, a^\dagger_i a^\dagger_j h^\dagger_k \ket{\mathrm{vac.}},
	\end{equation}
	as well as the corresponding energy $E_\mathrm{X^{-}}$. 
	Here $\ket{\mathrm{vac.}}$ is the vacuum state and $c_{kij}$ are numerically found expansion coefficients with $i>j$. In addition, we consider the single-electron states $\ket{l}=a_l^+\ket{\mathrm{vac.}}$ with the corresponding energies $E_{l}$. 
	
	\subsection{Optical spectra} 
	
	The radiative transitions in the many-particle system are described in the dipole approximation with the interband dipole moment operator~\cite{zielinski10}
	\begin{equation} \label{eq:tranistionop}
		\bm{D}=  \sum_{i,j} {\bm{d}}_{ij} \ h_i a_{j} \, .
	\end{equation}
	The matrix elements of the dipole moment between single-particle states are expressed in the tight-binding approach by 
	\begin{equation*}
		\bm{d}_{ij} \approx -e  \sum_{n}^{N_a} \sum_{\alpha}^{20} \varphi^{\mathrm{(v)}*}_{i,\alpha}(\bm{R}_{n}) \varphi^{\mathrm{(c)}}_{j,\alpha}(\bm{R}_{n}) \, \bm{R}_{n},
	\end{equation*}
	with the indices ``v" denoting a state from the valence band and ``c" a state from the conduction band. Here we neglect the contributions coming from the terms involving different nodes or orbitals,
	$\mel{\bm{R}_{m}; \,\alpha}{\bm{d}}{\bm{R}_{n}; \,\beta} \approx -e \, \bm{R}_n \, \delta_{mn}\delta_{\alpha\beta}$.
	
	The emission line intensity $I$ for the optical transition $\ket{X^-} \rightarrow \ket{l}$ is then given by
	\begin{align} \label{eq:oscstrength}
		I_l &= \frac{2 m_0}{\hbar^2 e^2} {\qty(E_\mathrm{X^{-}} - E_{l})} \sum_{\nu=x,y,z} {\abs{\mel{l}{D_{\nu}}{X^-}}^2},
	\end{align}
	where $m_0$ is the free electron mass
	and the matrix element $\mel{l}{D_{\nu}}{X^-}$ is related to the expansion coefficients for trion states (see Appendix).
	
	\section{Results}
	\subsection{QD spectra} \label{sec:results1}
	
	\begin{figure}[tb]
		\begin{center}
			\includegraphics[width=\columnwidth]{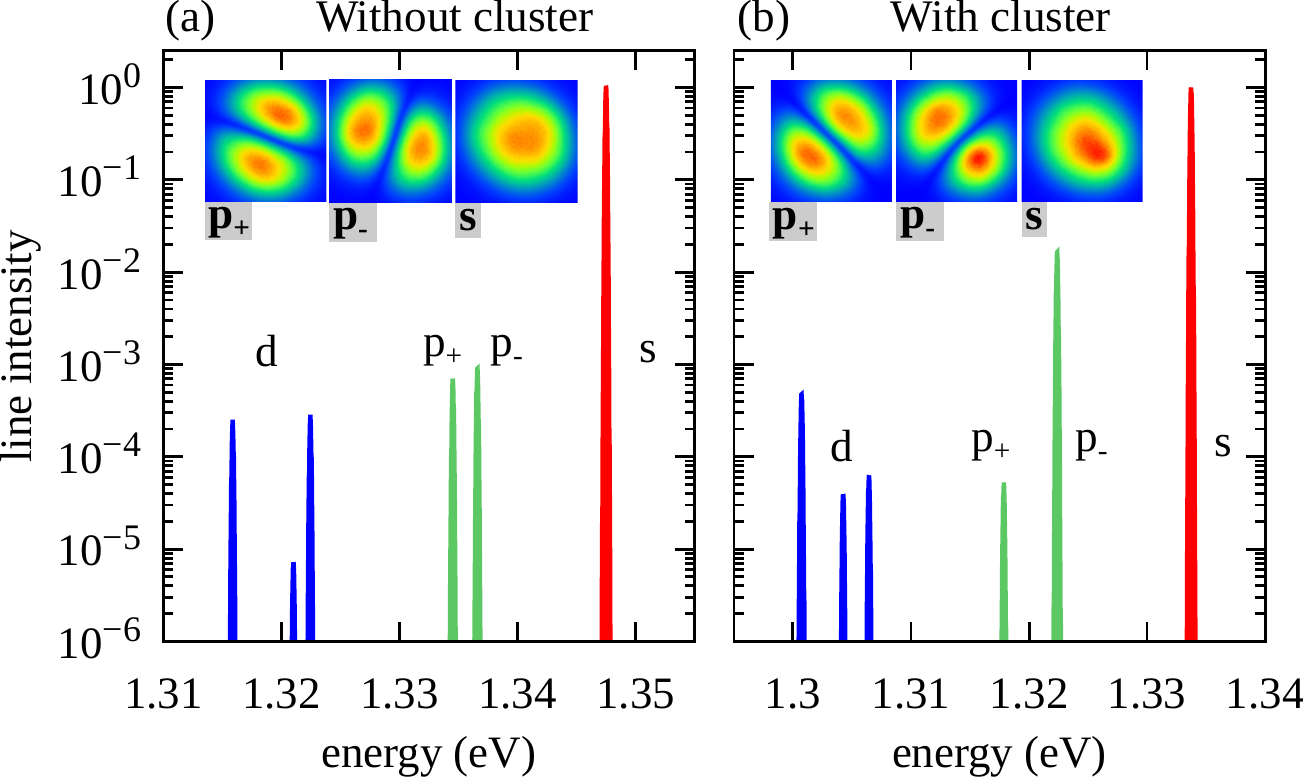}
		\end{center}
		\caption{ \label{fig:spectra_theo}
			Line intensity (logarithmic scale) of the transitions for a QD (a) without and (b) with a high In composition cluster as described in Sec.~\ref{sec:shape}, normalized to the $s$-shell transition. The widths of the lines are artificial. A small magnetic field was added to lift the spin degeneracy. The panels present probability densities of $s$ and $p$-type electron states in the (001) plane in the same crystallographic orientation as in Fig.~\ref{fig:comp}(c,d).} 
	\end{figure}
	
	We now use the tight binding model to calculate the optical spectra in the vicinity of the fundamental transition line as shown in Fig.~\ref{fig:spectra_theo} for a QD (a) without and (b) with composition cluster. For both cases we find a strong fundamental exciton transition line (red) at 1.348~meV in (a) and at 1.334~meV in (b), respectively. About 15~meV below we see a double peak structure, which corresponds to the radiative Auger transition with the remaining electron being promoted to the $p$ shell. Due to the in-plane asymmetry of the QD (including \blue{composition disorder,} strain and piezoelectric field \cite{Bester2005a,Zielinski2005a}) the $p$ shell is split into the $p_-$ and the $p_+$ line. In addition, about $30$~meV below the fundamental line, we find three Auger lines belonging to the $d$-shell transitions with different oscillator strengths.
	
	The general behaviour of the calculated spectra agrees well with the observed experimental data (cf. Fig.~\ref{fig:spectrum-exp}), in particular the energetic position of the lines is similar to the experiment. 
	However, without the cluster, we find that the intensity of the two $p$-shell lines is similar, which is in contrast to most of the experimental observations (except QD2) displaying a strong intensity difference between these lines.  Only if we include a cluster, the strong asymmetry between the two $p$-shell Auger lines is well reproduced.
	The role of the asymmetry can be understood when looking at the wave functions of the electronic states as shown in the insets in Fig.~\ref{fig:spectra_theo}. By adding an indium-rich cluster to the QD, the wave functions become distorted towards the cluster, making them asymmetric. In particular, for the $p$-shell wave function, the asymmetry affects the $p_-$ much more strongly than the $p_+$  wave function, thus leading to the different oscillator strengths. 
	\blue{We recall that even without the cluster, the exact symmetry of the system is already lowered due to the ``noise" in the composition profile resulting from the random choice of the atoms at the cation sites, as described in the Model section. However, the effect of such noise is averaged over the entire wavefunction of the QD. Therefore, the effect of random alloying is weaker than that of a large compositional cluster for the considered Auger transitions.}
	\begin{table}[tb]
		\centering
		\begin{ruledtabular}
			\begin{tabular}{c|ccc}
				& $p_- ~(10^{-3})$ & $p_+~(10^{-3})$ &ratio $p_+/p_-$  \\
				\hline
				QD1 &  1.2 &  0.34  & 0.28\\ 
				QD2 &  2.7 & 1.9  &0.70\\
				QD3 &  2.2 &  0.0  & 0.0 \\
				QD4 &  0.50 & 0.97  & 1.9\\ \hline  
				Without cluster &  \blue{0.95} & \blue{0.77} & \blue{0.82}\\ 
				High-In cluster &  17 & 0.06  & 0.004\\ 
				Low-In cluster &  \blue{2.1} & 0.13  & 0.06\\ 
			\end{tabular}
		\end{ruledtabular}
		\caption{    \label{tab:intensities}
			The intensities of the $p_-$ and $p_+$ Auger lines as well as their ratio $p_+/p_-$ at vanishing magnetic field relative to the fundamental transition for the four experimentally studied QDs (cf. Fig.~\ref{fig:bfield}) as well as for the theoretically modelled QDs, with and without an indium cluster (cf. Fig.~\ref{fig:spectra_theo}).}
	\end{table}
	
	Let us now have a closer look at the relative strength of the two $p$-shell Auger lines observed in the experiment, where the resonance fluorescence spectra of four different QDs labeled QD1 to QD4 were measured (see also Fig.~\ref{fig:bfield}). We have extracted the intensity of the $p$-shell Auger lines at zero magnetic field relative to the fundamental trion line in Tab.~\ref{tab:intensities}. The intensities are estimated by integration of the respective lines and  compared with the fundamental transition measured at shorter integration times. As QDs are not identical, we find a large variety of relative intensities of the two $p$-shell Auger lines ranging from $0.5\cdot 10^{-3}$ (QD4) to $2.7\cdot 10^{-3}$ (QD2) for the $p_-$-line and up to $1.9\cdot 10^{-3}$(QD2) for the $p_+$-line. We also find that the ratio $p_+/p_-$ between the two Auger lines differs greatly between the individual QDs, the extreme case of QD3 only showing a single $p_-$-line. In most cases we find that $p_-$ is much stronger than $p_+$. Only for QD4 this ratio is inverted, which can be traced back to a broad $p_+$ line, while the peak intensity of the $p_-$ is higher than $p_+$.
	
	We compare these values to our theoretical calculations, also displayed in Tab~\ref{tab:intensities}. 
	For the case without cluster we find the relative intensities of the two $p$ lines to be \blue{$0.95\cdot 10^{-3}$} and \blue{$0.77\cdot 10^{-3}$}. In contrast, for the case of a cluster with a high In concentration (as defined in Sec.~\ref{sec:model}), there is a pronounced difference in the line intensities, with the ratio $p_+/p_-$ being almost zero as for QD3. We also see that intensity of the lines can be strongly increased, when adding a cluster to the QD with the relative $p_-$ intensity reaching a value of $17\cdot 10^{-3}$. When the In concentration in the cluster becomes lower, the relative intensity of the $p_-$ line is reduced to \blue{$2.1\cdot 10^{-3}$}, which is very similar to the range of observed relative strengths in the experiment. The strength of the $p_+$ is less affected, resulting in a ratio of $p_+/p_-=0.06$, which is similar to the ratios found for QD1 and QD3, which indicates that the structure assumed in our computational model may roughly correspond to these particular samples.
	
	The spin-flip Auger transition, represented in Fig.~\ref{fig:SO-diagram}(c) may conserve the axial projection of the angular momentum upon transferring the angular momentum between the orbital and spin degrees of freedom. One could therefore expect that this transition may be allowed in the presence of spin-orbit couplings. Our simulations show, however, that this transition is one order of magnitude weaker than the spin-conserving one [Fig.~\ref{fig:SO-diagram}(b)] already in the structure without a cluster. Further symmetry breaking by including the cluster enhances the magnitude of the $p_-$ spin-conserving transition, while the intensity of the spin-flip one remains nearly unchanged. However, the spin-flip transitions can be important for the $p_+$ line in the presence of a cluster. As its strength is typically weaker, we conclude, therefore, that spin orbit effects are by far dominated by the effects of symmetry breaking in the radiative Auger process. When presenting the computational results, we always sum over the two spin configurations of the final state.
	
	We stress that both for the experimental and the theoretical line intensities a range of values is found. The overall agreement between the theory and experiment is very good, while tuning the form and magnitude of the symmetry breaking as well as the QD geometry may be required to reproduce the spectrum of a specific QD. Additional calculations as a function of the QD size (with a cluster which is also scaled) show a decreasing trend for the $p$-shell Auger transitions (in fact the dominant $p_-$ transition) as the QD becomes smaller.
	
	\subsection{Magnetic field dependence}
	
	\begin{figure*}[tb]
		\begin{center}
			\includegraphics[width=2\columnwidth]{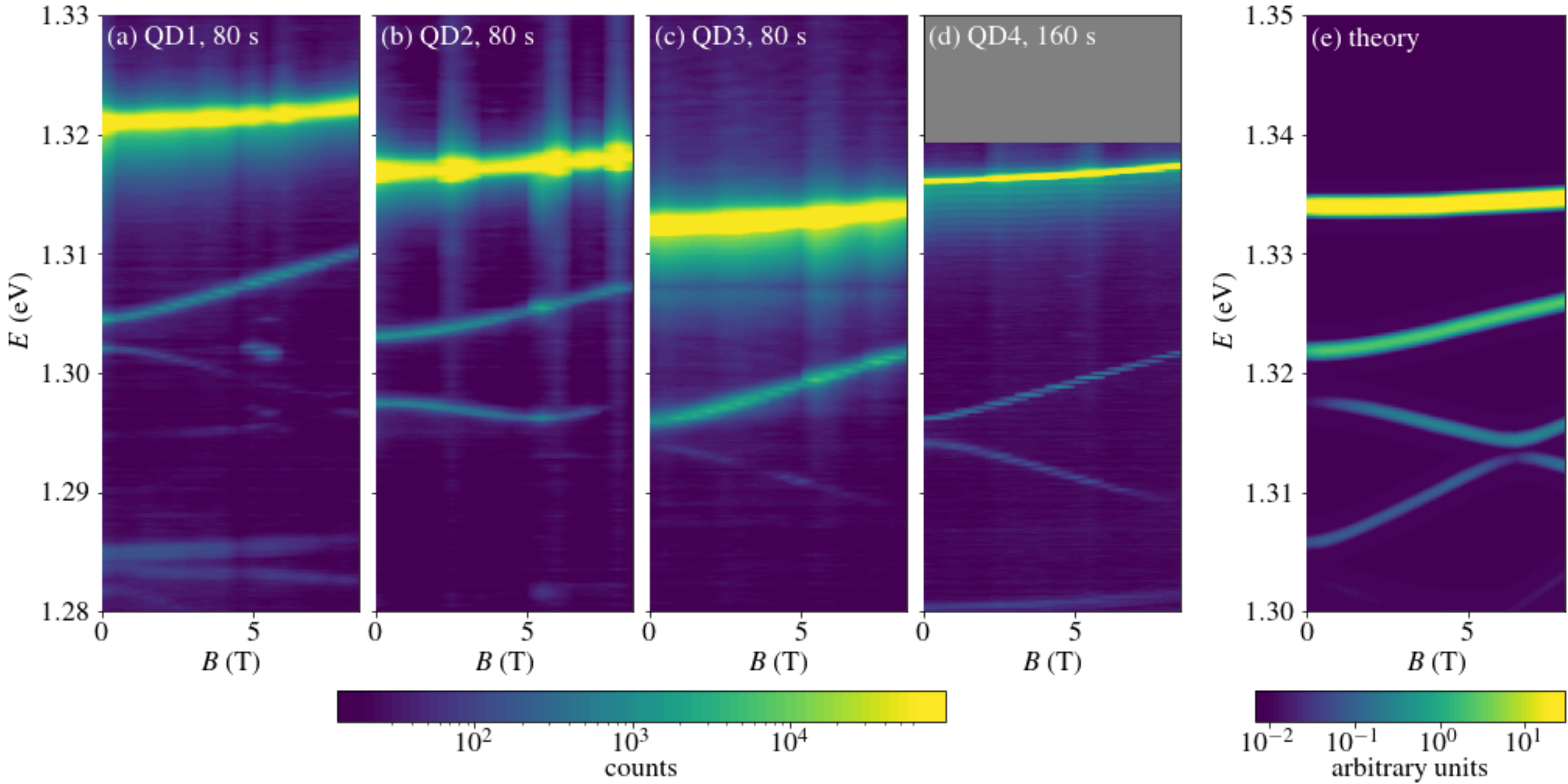}\\
		\end{center}
		\caption{\label{fig:bfield}
			Experimentally measured spectra of four different QDs (a-d) and theoretically calculated spectrum of a QD (e) as a function of the external magnetic field. The colorbar refers to the experimental plots (a-d) and shows the number of detection counts within the integration period indicated in each panel. \blue{In the case of the theoretical data, the color is related logarithmically to the oscillator strength (with an artificial broadening).}}
	\end{figure*}
	
	Another interesting feature of the Auger lines is their dependence on an applied external magnetic field. Examples of the measured magnetic field dependence are shown in Fig.~\ref{fig:bfield}(a-d) \cite{Lobl2020}. While the fundamental transition remains rather unchanged under the influence of a magnetic field, apart from a slight diamagnetic shift, the $p$-shell Auger lines show a typical Zeeman shift behaviour. In some cases one observes an anti-crossing with the $d$-shell Auger-lines [e.g. at a magnetic field of $B=6$~T in Fig.~\ref{fig:bfield}(b)]. The overall behaviour of the four different dots QD1 to QD4 is rather similar with the main difference found in the line intensities as discussed in the previous section. \blue{However, the visibility of the $d$-shell varies from QD to QD. This could be a consequence of the proximity of the continuum states in the wetting layer. 
	It is known that the thickness of the wetting layer and therefore the energetic distance between wetting layer and QD levels fluctuates~\cite{Bart2022}.}
	
	In the theoretical modeling we include the interaction with the magnetic field in the tight-binding approach by the Peierls substitution and on-site Zeeman terms~\cite{Graf1995,Boykin2001}. A detailed description of the model is given in Ref.~\cite{Gawarecki2020}.
	The obtained spectra are shown in Fig.~\ref{fig:bfield}(e) for a QD with a high-concentration indium cluster. We find that the overall agreement with the experimental data is very good. In particular, the energetic behaviour of the lines can be well reproduced and we also find the anticrossing between the different shells. We can also reproduce the intensity behaviour, in particular the difference in the two $p$-shell lines, consistently for non-zero magnetic fields. The theoretical result, as well as the measurement shown in Fig.~\ref{fig:bfield}(c), show an enhancement of the weaker ($p_+$) radiative Auger line relative to the $p_-$. This effect of equalizing the intensities of the two $p$-shell branches can be attributed to partial restoration of the symmetry of envelope functions by the magnetic field. 

	\section{Group theoretical considerations}	
	
	The appearance of the radiative Auger lines by the symmetry breaking can be explained using group theory. While these arguments are not sufficient to get the strength of the lines as obtained from the numerical calculation, they allow one to predict some qualitative properties of the Auger process for various system symmetries. 
	
	Focusing on a major symmetry breaking on the level of the QD shape, in the subsequent analysis we do not consider atomic disorder, \blue{band mixing}, spin-orbit coupling nor the effect of a magnetic field. In such an approximation, optical transitions take place only if the envelopes of the initial and the final state belong to the same irreducible representation of a given group. The participating states in the optical transition are the electron states $\vert l \rangle$ and the ground trion state $\ket{X^-}$, for which we investigate the representations linked to their envelopes. 
	In this section it is convenient to highlight the symmetry of the single-particle states $|l\rangle$ by using shell-based notation, where  we label the states with their approximate envelope type starting with $l = s$ followed by the lower and upper $p$-shell states $l= p_-$ and $p_+$ and then the $d$-shell states.
	
	\begin{table}[tb]
		\begin{ruledtabular}
			\begin{tabular}{c|ccccc}
				$C_{2v}$ & $E$ & $C_{2}$ & $\sigma_{v}(110)$ & $\sigma_{v}(\bar{1}10)$ & basis functions \\ 
				\hline \\ [-1.5ex]
				$\Gamma_1$ & 1 & 1 & 1 & 1 & 1, $z$, $x^2$, $y^2$, $z^2$ \\
				$\Gamma_2$ & 1 & 1 & -1 & -1  & $xy$ \\
				$\Gamma_3$ & 1 & -1 & 1 & -1  & $y$, $yz$ \\
				$\Gamma_4$ & 1 & -1 & -1 & 1   & $x$, $xz$\\
				\hline \\ [-1.5em]
				$C_{s}$ & $E$ &&$\sigma_{v}(110)$ && basis functions \\ 
				\hline \\ [-1.5ex]
				$\Gamma_1$ & 1 &&  1 && 1, $y$, $z$ \\
				$\Gamma_2$ & 1 && -1 && $x$ 			
			\end{tabular} 
		\end{ruledtabular}
		\caption{\label{tab:C2v} Character table of the irreducible representations of the $C_{2v}$ and the the $C_{s}$ point group \cite{Bradley1972}. The $E$ is an identity element, $\sigma$ are reflection planes described by the Miller indices, $C_2$ is the rotation by $\pi$. Note, that in the basis functions here, in contrast to the other parts of the article, $x$ direction is $[110]$ and $y$ is [$\bar{1}10$]. } 
	\end{table} 
	
	The bare elliptical QD [Figs.~\ref{fig:comp}(a) and ~\ref{fig:comp}(c)] is described by the $C_{2v}$ group, which contains one operation of rotation and two reflection planes (see the character Table~\ref{tab:C2v}). Due to its transformation rules, the electron ground state $\ket{s}$ belongs to the identity representation $\Gamma_1$. On the other hand, the first excited states $\ket{p_-}$ and $\ket{p_+}$ belong to $\Gamma_3$ and $\Gamma_4$, respectively [which can also be seen in the insets in Fig.~\ref{fig:spectra_theo}(a)]. Being a ground state, $\ket{X^-}$ belongs to the $\Gamma_1$ representation. In consequence, the fundamental transition from $\ket{X^-}$ to the $\ket{s}$ electron state is allowed, while the Auger transitions to $\ket{p_-}$ and $\ket{p_+}$ are prohibited by group theoretical arguments.
	
	In the case of the QD with composition cluster [Figs.~\ref{fig:comp}(b) and ~\ref{fig:comp}(d)], the symmetry is reduced to the $C_{s}$ group, which contains a reflection by a single plane (see the character Table~\ref{tab:C2v}). As before, trion $\ket{X^-}$ and electron $\ket{s}$ are linked to the identity representation $\Gamma_1$. However, due to the symmetry reduction by the cluster, $\ket{p_-}$ and $\ket{p_+}$ now belong to the $\Gamma_1$ and $\Gamma_2$, respectively. In consequence, the Auger transition $\ket{X^-} \leftrightarrow \ket{p_{-}}$ is now allowed. This is in agreement with the observation of a large peak related to this process, as observed in Fig.~\ref{fig:spectra_theo}(b) and in the experimental data. The symmetry reduction from $C_{2v}$ to $C_{s}$ was also shown to be crucial for optical activity of dark excitons in self-assembled QDs~\cite{Zielinski2015PRB}.
	
	We emphasize, that in an exact approach, due to random atomic disorder, all the considered systems have strictly $C_1$ symmetry, where all of the states resides in the same representation. This weakly allows even further Auger transitions. Additionally, spin-orbit coupling mixes the spin and shell configurations. 
	Typically, such effects have a smaller impact compared to the strong mesoscopic effects related to structural shape anisotropy, which is in agreement with our simulations in Fig.~\ref{fig:spectra_theo} displaying a strong $p_-$ and a much weaker $p_+$ line.
	
	
	\section{Conclusions}
	\label{sec:concl}
	
	In conclusion, we have presented a tight-binding approach to model the radiative Auger lines as observed recently in semiconductor QDs. We were able to reproduce the line positions and their behaviour in a magnetic field. We showed that in the presence of an additional symmetry breaking, induced for example by a cluster with increased indium content, the relative strength of the radiative Auger lines agrees with the experimentally observed values. Other mechanisms of symmetry breaking might lead to similar effects. However, the apparently most natural explanation invoking charged traps does not seem plausible: One would require deep traps (to remain charged in the Coulomb blockade regime of the experiment) at a sufficiently high density to  provide a charge sufficiently close to a statistically substantial fraction of QDs. This seems to contradict the general knowledge on the high-quality QD samples. 
	
	Additionally, we linked the Auger lines to group theoretical considerations that highlight the importance of symmetry breaking below the $C_{2v}$ point group, characterizing elliptical QDs on the mesoscopic level. 
	
	Our results give new insight in the interplay of structural symmetry and the emergence of radiative Auger lines in QDs. By relating the appearance of these lines to the symmetry-breaking features in the QD structure we show that spectroscopic measurements can reveal information on the morphology of the QD.
	
	\acknowledgements
	We thank T. K. Bracht for help with the figures. \blue{We are also grateful to Micha{\l} Gawe{\l}czyk for sharing his implementation of the blur algorithm and to Micha{\l} Zieliński for discussions.} CS, LZ, GNN, MCL, and RJW acknowledge financial support from Swiss National Science Foundation project 200020\_204069 and NCCR QSIT. LZ and GNN acknowledge support from the European Union’s Horizon 2020 Research and Innovation Programme under the Marie Sk\l odowska-Curie grant agreement numbers 721394 / 4PHOTON (LZ), 861097 / QUDOT-TECH (GNN). KG and PM acknowledge funding from the Polish National Science Centre (NCN) under Grant No. 2016/23/G/ST3/04324.

	\appendix
	
	\section{Model implementation}\label{sec:app_model}
	
	Here we present some additional details of our model complementing Sec.~\ref{sec:model}, including explicit expressions for the Coulomb matrix elements. 
	
	\blue{The piezoelectric potential [$V_\mathrm{p}(\rr)$] is calculated by solving the Poisson-like equation
		\begin{equation}
			\label{eq:Poisson}
			\epsilon_{0} \div{\qty[\epsilon_r(\rr) \grad{V_\mathrm{p}(\rr)}]} = - \div{\bm{P}}, 
		\end{equation}
		where
		$\epsilon_0$ is the vacuum permitivity, $\epsilon_r(\rr)$ is the position-dependent relative permittivity, and $\bm{P}$ is the polarization, which is calculated from the local strain tensor elements~\cite{bester06b,Caro2015}
		\begin{align*}
			\bm{P}  & =   2 e_{14} \begin{pmatrix} \epsilon_{yz} \\  \epsilon_{xz} \\  \epsilon_{xy} \end{pmatrix}  + 2B_{114} \begin{pmatrix} \epsilon_{xx} \epsilon_{yz} \\ \epsilon_{yy} \epsilon_{xz} \\ \epsilon_{zz} \epsilon_{xy} \end{pmatrix} + 4B_{156} \begin{pmatrix} \epsilon_{xz} \epsilon_{xy} \\ \epsilon_{yz} \epsilon_{xy} \\ \epsilon_{yz} \epsilon_{xz} \end{pmatrix} \\ &
			+ 2B_{124} \begin{pmatrix} (\epsilon_{yy} + \epsilon_{zz} ) \epsilon_{yz} \\ (\epsilon_{xx} + \epsilon_{zz} ) \epsilon_{xz} \\ (\epsilon_{xx} + \epsilon_{yy} ) \epsilon_{xy} \end{pmatrix}.
		\end{align*}
		The strain tensor elements at cations are calculated following Ref.~\cite{pryor98b}, and at anions they are obtained by averaging the values from the neighboring cations. We take $\epsilon(\mathrm{InAs}) = 14.6$ and $\epsilon(\mathrm{GaAs}) = 12.4$~\cite{Adachi1982,Blakemore1982}.
		The values of $e_{14}$, $B_{114}$, $B_{156}$, $B_{124}$ are taken from Ref.~\cite{bester06b}. We discretize Eq.~(\ref{eq:Poisson}) using the finite difference scheme for the atomistic grid~\cite{Pryor2015}. For the anion at position $\bm{R}_i$, this gives
		\begin{equation}
			\label{eq:discrA}
			\begin{split}	
				& \frac{4 \epsilon_0 }{a} \sum_{j}^{\mathrm{NN}(i)} \qty[\epsilon_r(\bm{R}_i + \bm{d_j}) + \epsilon_r(\bm{R}_i) ] \qty[V_\mathrm{p}(\bm{R}_i + \bm{d_j}) - V_\mathrm{p}(\bm{R}_i) ] = \\ &- \Big [ P_x(\bm{R}_i + \bm{d}_1) - P_x(\bm{R}_i + \bm{d}_2) - P_x(\bm{R}_i + \bm{d}_3)  \\ & + P_x(\bm{R}_i + \bm{d}_4)\Big ] -\ \Big [ P_y(\bm{R}_i + \bm{d}_1) - P_y(\bm{R}_i + \bm{d}_2) \\ & + P_y(\bm{R}_i + \bm{d}_3) - P_y(\bm{R}_i + \bm{d}_4)\Big ]
				-\Big [ P_z(\bm{R}_i + \bm{d}_1) \\ & + P_z(\bm{R}_i + \bm{d}_2) - P_z(\bm{R}_i + \bm{d}_3) - P_z(\bm{R}_i + \bm{d}_4)\Big ],
			\end{split}
		\end{equation}
		where NN$(i)$ denote the nearest neighbors of the $i$-th atom. For the unstrained lattice the positions of the surrounding cations are given by
		\begin{align*}
			\bm{d}_1 &= \frac{a}{4} (1,1,1), \\
			\bm{d}_2 &= \frac{a}{4} (-1,-1,1),\\
			\bm{d}_3 &= \frac{a}{4} (-1,1,-1), \\
			\bm{d}_4 &= \frac{a}{4} (1,-1,-1). \\
		\end{align*}
		In the case of the cation at $\bm{R}_i$, the discretization takes the form
		\begin{equation}
			\label{eq:discrC}
			\begin{split}	
				& \frac{4 \epsilon_0 }{a} \sum_{j}^{\mathrm{NN}(i)} \qty[\epsilon(\bm{R}_i - \bm{d_j}) + \epsilon(\bm{R}_i) ] \qty[V_\mathrm{p}(\bm{R}_i - \bm{d_j}) - V_\mathrm{p}(\bm{R}_i) ] = \\ & \Big [ P_x(\bm{R}_i - \bm{d}_1) - P_x(\bm{R}_i - \bm{d}_2) - P_x(\bm{R}_i - \bm{d}_3)  \\ & + P_x(\bm{R}_i - \bm{d}_4)\Big ] +\ \Big [ P_y(\bm{R}_i - \bm{d}_1) - P_y(\bm{R}_i - \bm{d}_2) \\ & + P_y(\bm{R}_i - \bm{d}_3) - P_y(\bm{R}_i - \bm{d}_4)\Big ]
				+ \Big [ P_z(\bm{R}_i - \bm{d}_1) \\ & + P_z(\bm{R}_i - \bm{d}_2) - P_z(\bm{R}_i - \bm{d}_3) - P_z(\bm{R}_i - \bm{d}_4)\Big ].
			\end{split}
		\end{equation}
		One should note that Eqs.~(\ref{eq:discrA}),~(\ref{eq:discrC}) are derived for the unstrained zinc-blende lattice. 
		We use this approach here with the relaxed atomic positions. However, this approximation can be justified by the fact that strain is typically on the order of a few percent. The resulting set of linear equations is solved numerically using the library PETSC~\cite{petsc}.
	}

	The starting point \blue{of further calculations} are the single particle wave functions $\ket{\Psi_{i}}$ as given in Sec.~\ref{sec:coulomb}, which are superpositions of the electronic wave functions of the atoms in the system. As the spin is taken into account, the basis for the sp$^3$d$^5$s$^*$ model contains $20$ atomic states.
	
	In the next step, we use the configuration-interaction (CI) Hamiltonian as given in Eq.~\eqref{eq:CI}. 
	The calculations at $B=0.1$~T were performed with the basis of $n_\mathrm{e} = 20$ and $n_\mathrm{h} = 20$.
	Due to a high computational cost of simulations, the results for magnetic-field dependence [Fig.~\ref{fig:bfield}(e)] were obtained in a reduced basis of $n_\mathrm{e} = 12$ and $n_\mathrm{h} = 12$. The difference between the p-shell Auger intensities obtained with $12$ and $20$ basis states for the QD without cluster and with low-In cluster is $30\%$ to $50\%$. On the other hand, with high-In cluster, these results are almost identical.
	In the tight-binding basis, the Coulomb matrix elements between a particle A and B are given by~\cite{Schulz2006,zielinski10}
	
	\begin{align*}
		V^{AB}_{i j j' i'} & \approx V_0 + \frac{\abs{e}^{2}}{4 \pi \epsilon_{0} \epsilon_{r}} \sum_{n,m \neq n}^{N_a} \sum_{\alpha,\beta}^{20} \\ & \times   \frac{\varphi^{(A)*}_{i,\alpha}(\bm{R}_{n}) \varphi^{(B)*}_{j,\beta}(\bm{R}_{m}) \varphi^{(B)}_{j',\beta}(\bm{R}_{m}) \varphi^{(A)}_{i',\alpha}(\bm{R}_{n})}{\abs{\bm{R}_{n} - \bm{R}_{m}}},\\
		V^{AB,\mathrm{exch}}_{i j i' j'} & \approx V^\mathrm{exch}_0 + \frac{\abs{e}^{2}}{4 \pi \epsilon_{0} \epsilon_{r}} \sum_{n,m \neq n}^{N_a} \sum_{\alpha,\beta}^{20} \\ & \times   \frac{\varphi^{(A)*}_{i,\alpha}(\bm{R}_{n}) \varphi^{(B)*}_{j,\beta}(\bm{R}_{m}) \varphi^{(A)}_{i',\beta}(\bm{R}_{m}) \varphi^{(B)}_{j',\alpha}(\bm{R}_{n})}{\abs{\bm{R}_{n} - \bm{R}_{m}}},
	\end{align*}
	where we took the value for GaAs $\epsilon_{r} = 12.4$, $V_0$ and $V^{\mathrm{exch}}_0$ account for the short-range on-site contributions, which we neglect here. \blue{Such elements could be calculated using the basis of atomic orbitals (e.g. the Slater orbitals)~\cite{zielinski10,Lee2001}. However, for direct-bandgap QDs, the on-site elements vanish much faster with an increasing QD size than the monopole-monopole long-range terms~\cite{Franceschetti1998,Luo2009,Zielinski2013}. For the InAs quantum dots considered in Ref.~\cite{Zielinski2013}, the on-site terms contributed only about 1\% to the direct Coulomb attraction for the ground electron-hole states, and 20\% to the exchange-interaction-induced splitting.} We calculate the \blue{two-center contributions} using the expansion according to the Fourier theorem
	$$
	\frac{1}{\abs{\bm{R}_{n} - \bm{R}_{m}}} = \frac{1}{(2 \pi)^3} \int \frac{4\pi}{q^2} e^{i \bm{q} (\bm{R}_{n} - \bm{R}_{m})} \dd {\bm{q}},
	$$
	which allows us to write 
	\begin{align}
		V^{AB}_{i j j' i'} \approx & \frac{\abs{e}^{2}}{8 \pi^3 \epsilon_{0} \epsilon_{r}} \int \frac{F^{(AA)}_{ii'}(\bm{q}) F^{(BB)*}_{j'j}(\bm{q})}{q^2}  \dd {\bm{q}}  \nonumber,\\
		V^{AB,\mathrm{exch}}_{i j  i'j'} \approx & \frac{\abs{e}^{2}}{8 \pi^3 \epsilon_{0} \epsilon_{r}} \int \frac{F^{(AB)}_{ij'}(\bm{q}) F^{(AB)*}_{i'j}(\bm{q})}{q^2}  \dd {\bm{q}}  \nonumber,
	\end{align}
	where 
	$$
	F^{(AB)}_{ij}(\bm{q}) = \sum_{n}^{N_\mathrm{a}} \sum_{\alpha}^{20} \varphi^{(A)*}_{i,\alpha}(\bm{R}_{n}) \varphi^{(B)}_{j,\alpha}(\bm{R}_{n}) e^{i \bm{q} \bm{R}_{n}},
	$$
	which we calculated efficiently using FINUFFT library~\cite{Barnett2019,Barnett2021,Lee2005}. \blue{To perform the calculations for $V^{AB}_{i j j' i'}$ and $V^{AB,\mathrm{exch}}_{i j  i'j'}$ we use spherical coordinates, where the $1/q^2$ singularity is removed by the Jacobian. We take the maximal value of $q$ in the integration as $\pi/a$.} \blue{We checked that the exchange-interaction-terms have a small impact on the considered Auger transitions; they are included in the model mainly for completeness.}
	After the diagonalization in the basis of $n_\mathrm{e}$ electron and $n_\mathrm{h}$ hole states, we obtain the energies
	$E_\mathrm{X^{-}}$ and coefficients $c_{kij}$ from which the trion ground state $\ket{X^-}$ is formed as described in Eq.~\eqref{eq:trion}.

	The expansion coefficients of the Coulomb-coupled trion states on the other hand enter in the calculation of the 
	matrix element of the transition operator $\mathbf{D}$ given in Eq.~\eqref{eq:tranistionop},
	\begin{align*}
		\mel{l}{\bm{D}}{X^-} &=   \bra{\mathrm{vac.}} a_l  \sum_{i',j'} \bm{d}_{i'j'} \ h_{i'} a_{j'}  \sum_{\substack{k,i,j }} c_{kij} \\ & \times a^\dagger_i a^\dagger_j h^\dagger_k \ket{\mathrm{vac.}} \\& = \sum_{k,i,j} c_{kij} \qty( \bm{d}_{ki} \delta_{lj} - \bm{d}_{kj} \delta_{li}),
	\end{align*}
	where $i > j$.  

	\bibliography{library.bib,doris.bib,krzysiek.bib}

\end{document}